\documentclass[12pt]{iopart}

\begin{document}
\newcommand{\goo}{\,\raisebox{-.5ex}{$\stackrel{>}{\scriptstyle\sim}$}\,}
\newcommand{\loo}{\,\raisebox{-.5ex}{$\stackrel{<}{\scriptstyle\sim}$}\,}

\title{Evaluation of the hyperon binding energy via statistical production of hypernuclei}

\author{A.S.~Botvina$^{1,2}$, M.~Bleicher$^1$, N.~Buyukcizmeci$^3$}

\address{$^1$Frankfurt Institute for Advanced Studies and ITP J.W. Goethe University, D-60438 Frankfurt am Main, Germany\\
$^2$Institute for Nuclear Research, Russian Academy of Sciences, 117312 Moscow, Russia\\
$^3$Department of Physics, Selcuk University, 42079 Campus, Konya, Turkey}
\ead{nihal@selcuk.edu.tr}
\vspace{10pt}
\begin{indented}
\item[] Received: 29 November 2017
\end{indented}

\begin{abstract}
In nuclear reactions of high energy one can simultaneously produce a lot 
of hypernuclei after the capture of hyperons by nuclear residues. We 
consider statistical disintegration of such hypernuclear systems and the 
connection of fragment production with the binding energies of hyperons. 
It is demonstrated that the hyperon binding energies can be effectively 
evaluated from the yields of different isotopes of  hypernuclei. 
The double ratio method is suggested for this purpose. The advantage of 
this procedure is its universality and the possibility to involve many 
different isotopes. This method can also be applied for multi-strange 
nuclei, which binding energies were very difficult to measure in 
previous hypernuclear experiments. 
\end{abstract}

\submitto{\JPG}
%\maketitle

\section{Introduction}
A promising way to produce hypernuclei is to use the copious production of 
hyperons ($\Lambda,\Sigma,\Xi,\Omega$) in relativistic nuclear reactions 
with their subsequent capture by nuclei. 
Hypernuclei live significantly longer than the typical reaction times. 
Baryons with strangeness embedded in the nuclear 
environment allow for approaching the many-body aspect of the strong 
three-flavor interaction (i.e., including $u$, $d$, and $s$ quarks) at 
low energies. Also hypernuclei 
can serve as a  tool to study the hyperon--nucleon and hyperon--hyperon 
interactions. The investigation of reactions leading to hypernuclei and 
the structure of hypernuclei is the progressing 
field of nuclear physics, since it provides complementary methods to 
improve traditional nuclear studies and open new horizons for studying 
particle physics and nuclear astrophysics 
(see, e.g., \cite{Ban90,Sch93,Has06,Gal12,Buy13,Hel14} 
and references therein). 

We emphasize specially a possibility to form hypernuclei in the deep-inelastic 
reactions leading to fragmentation processes, as they were discovered 
long ago \cite{Dan53}. 
Many experimental collaborations STAR at RHIC \cite{star}, ALICE at LHC 
\cite{alice}, CBM \cite{Vas17}, 
HypHI, Super-FRS, R3B at FAIR \cite{saito-new,super-frs}, BM@N, MPD 
at NICA \cite{nica}) plan to investigate hypernuclei and 
their properties in reactions induced by relativistic hadrons and ions. 
The limits in isospin space, particle unstable states, multiple strange 
nuclei and precision lifetime measurements are unique topics of these 
fragmentation reactions. A capture of hyperons by large nuclear residues 
formed in peripheral collisions is also interesting since it provides a 
natural way to study large bulbs of hypermatter and its evolution, 
for example, the liquid-gas type phase transition. It was theoretically 
demonstrated \cite{Buy13,Bot07,Das10,Bot11,Bot13,Bot15,Bot17,Cas95,giessen} 
that in such a way it is possible to produce all kind of hypernuclei with a 
very broad isospin content. 
There were also experimental confirmations of such processes leading 
at least to single hypernuclei \cite{Arm93,Ohm97,saito-new}. 
In addition, complex multi-hypernuclear systems incorporating more than 
two hyperons can be created in the energetic nucleus-nucleus collisions 
\cite{Bot11,Bot17}. 
This may be the only conceivable method to go beyond double hypernuclei, and 
obtain new experimental information on properties of multi-hyperon systems.
In this Rapid Communication we demonstrate how the important knowledge on 
the hyperon binding energies, including in multi-strange nuclei, can be 
extracted from analysis of the relative yields of hypernuclei. 

In high energy nucleus-nucleus and hadron-nucleus collisions the production 
of strangeness correlates with particle production, therefore, emission of 
many nucleons can accompany the production of hyperons. An initial nucleus 
can loose many nucleons, and, as known from normal nucleus interactions, 
these processes are leading to a high excitations of remaining residual 
nuclei, see e.g. Refs.~\cite{SMM,Xi97,turzo}. 
In this case the capture of a produced hyperon will be also realized 
at the excited nuclei. As a result such deep-inelastic processes can 
form large hyper-residues with very broad distribution in 
mass and excitation energy. It was demonstrated in our previous works 
\cite{Bot13,Bot15,Bot17} that the yields of the hypernuclear residues in 
peripheral ion collisions will saturate with energies above 3--5 A GeV(in the laboratory frame).

The reactions of formation of excited nuclear residues in high-energy 
nucleus-nucleus and hadron-nucleus collisions were intensively studied 
in connection with fragmentation and multifragmentation processes. 
In particular, masses and excitation energies of the residues are known 
from experimental and theoretical works, e.g., Refs.~\cite{Xi97,Bot17}. 
At high excitation energy the dominating decay mode is a multifragmentation 
process \cite{SMM,Poc97,EOS}. The hyperon interactions in a nucleus are 
similar to nuclear ones, and its potential is around 2/3 of the nucleon 
one. Therefore, we believe, that an addition of few hyperons to a 
multi-nucleon system can not change its disintegration behavior. 
According to the present understanding, multifragmentation is a 
relatively fast process, with a characteristic time around 100 fm/c, 
where, nevertheless, a high degree of equilibration (chemical equilibrium) 
is reached. This is a consequence of the strong interaction between 
baryons, which are in the vicinity of each other in the freeze-out volume. 

The statistical models have demonstrated very good agreement in comparison 
with fragmentation and multifragmentation data \cite{SMM,Xi97,EOS,Ogu11}. 
It is naturally to extend the statistical approach for hypernuclear systems. 
The same numerical methods used previously for 
execution of the models can be extended. The statistical multifragmentation 
model (SMM), which was very 
successfully applied for description of normal multifragmentation 
processes, was generalized for hypernuclei in Ref.~\cite{Bot07}. 
The break-up channels are generated according to their statistical weight. 
The Grand Canonical approximations leads to the following average yields of 
individual fragments with the mass (baryon) number $A$, charge $Z$, and the 
$\Lambda$-hyperon number $H$: 
\begin{eqnarray} \label{yazh} 
Y_{\rm A,Z,H}=g_{\rm A,Z,H}\cdot V_f\frac{A^{3/2}}{\lambda_T^3} 
{\rm exp}\left[-\frac{1}{T}\left(F_{A,Z,H}-\mu_{AZH}\right)\right], 
\nonumber\\ 
%\mu_{AZH}=A\mu+Z\nu+H\xi~, 
\mu_{AZH}=A\mu+Z\nu+H\xi .~~~~
\end{eqnarray} 
Here $T$ is the temperature, $F_{A,Z,H}$ is the internal free energies of 
these fragments, $V_f$ 
is the free volume available for the translation motion of the fragments, 
$g_{\rm A,Z,H}$ is the spin degeneracy factor of species 
$(A,Z,H)$, $\lambda_T=\left(2\pi\hbar^2/m_NT\right)^{1/2}$ is the 
baryon thermal wavelength, $m_N$ is the average 
baryon mass. The chemical potentials $\mu$, $\nu$, and $\xi$ are 
responsible for the mass (baryon) number, charge, and strangeness 
conservation in the system, and they can be numerically found from the 
corresponding conservation laws. In this model the statistical ensemble 
includes all break-up channels composed of baryons and excited 
fragments. The primary fragments are formed in the freeze-out volume $V$. 
We use the excluded volume approximation $V=V_0+V_f$, where 
$V_0=A_0/\rho_0$ ($A_0$ is the total baryon number and 
$\rho_0\approx$0.15 fm$^{-3}$ is the normal nuclear 
density), and parametrize the free volume $V_f=\kappa V_0$, with 
$\kappa \approx 2$, as taken in description of experiments 
in Refs.~\cite{Xi97,EOS,Ogu11}.

\section{Double ratio method for hypernuclei} 

The following model development depends on the physical processes which 
are the most adequate to the analyzed reactions. 
For example, nuclear clusters in the freeze-out volume can be described in 
the liquid-drop approximation: Light fragments are treated as elementary 
particles with corresponding spins and translational degrees of freedom 
("nuclear gas"). Their binding energies were taken from experimental data
\cite{Ban90,Has06,SMM}. Large fragments are treated as heated liquid drops. 
In this way one can study the nuclear liquid-gas coexistence of hypermatter 
in the freeze-out volume. The internal free energies of these fragments are 
parametrized as the sum of the bulk ($F_{A}^B$), the surface 
($F_{A}^S$), the symmetry ($F_{AZH}^{\rm sym}$), the Coulomb 
($F_{AZ}^C$), and the hyper energy ($F_{AH}^{\rm hyp}$): 
\begin{equation}  \label{fld}
F_{A,Z,H}=F_{A}^B+F_{A}^S+F_{AZH}^{\rm sym}+F_{AZ}^C+F_{AH}^{\rm hyp}~~.
\end{equation}
One can find in Refs.~\cite{Buy13,Bot07,SMM} all details of this approach. 
In Ref.~\cite{Bot07} we have suggested 
that the hyper term $F_{AH}^{\rm hyp}$ is determined only by the binding 
energy of hyperons with the following parametrization:
\begin{equation}
F_{AH}^{\rm hyp}=(H/A)\cdot(-10.68 A + 21.27 A^{2/3}) MeV , 
\end{equation}
where the binding of hypernuclei is proportional to the fraction of hyperons
in matter ($H/A$). 
As was demonstrated in Refs.~\cite{Bot07,Buy13} this parametrization of 
the hyperon binding energy describes available experimental data quite 
well. It is important that two boundary physical effects are 
correctly reproduced: The binding energies of light hypernuclei (if a hyperon 
substitutes a neutron) can be lower than in normal nuclei, since the 
hyperon-nucleon potential is smaller than the nucleon-nucleon one. However, 
since the hyperon can take the lowest s-state, it can increase the nuclear 
binding energies, specially for large nuclei. Within the SMM approach we have 
performed an analysis of fragment and hyper-fragment production from excited 
hypernuclear systems. A transition from the compound hyper-nucleus to the 
multifragmentation regime was under investigation too \cite{Bot07,Buy13}. 

It is convenient to rewrite the above formulas in order to show separately 
the binding energy $E_{A}^{\rm bh}$ of one hyperon at the temperature $T$ 
inside a hypernucleus with $A,Z,H$~: 
\begin{equation} \label{hybi}
E_{A}^{\rm bh} = F_{A,Z,H} - F_{A-1,Z,H-1} ~. 
\end{equation}
Since $\Lambda$-hyperon is usually bound, this value is negative. 
Then the yield of hypernuclei with an additional $\Lambda$ hyperon can be 
recursively written by using the former yields: 
\begin{equation} \label{yazhy} 
Y_{\rm A,Z,H}=Y_{\rm A-1,Z,H-1} \cdot C_{\rm A,Z,H} \cdot 
{\rm exp}\left[-\frac{1}{T}\left(E_{A}^{\rm bh}-\mu-\xi\right)\right], 
\end{equation} 
where 
$C_{\rm A,Z,H}=(g_{\rm A,Z,H}/g_{\rm A-1,Z,H-1}) \cdot (A^{3/2}/(A-1)^{3/2})$ 
depends mainly on the ratio of the spin factors of $A,Z,H$ and 
$A-1,Z,H-1$ nuclei, and very weakly (especially for large nuclei) on $A$. 
Since in the liquid-drop approximation we assume that the fragments with 
$A > 4$ are 
excited and do populate many states (above the ground) according to the given 
temperature dependence of the free energy, then we take $g_{\rm A,Z,H} = 1$. 
Within SMM we can connect the relative yields of hypernuclei with the 
hyperon binding energies. It is interesting that in this formulation 
one can use other parametrizations to describe nuclei in the freeze-out. This 
statistical approach is quite universal, and only small corrections, like 
the table-known spins and energies, may be required for more extensive 
consideration. 

We suggest the following receipt for obtaining information on the 
binding energies of hyperons inside nuclei. Let us take two hyper-nuclei 
with different masses, ($A_1,Z_1,H$) and ($A_2,Z_2,H$), 
together with nuclei which differ from them only by one $\Lambda$ hyperon. 
When we consider the double ratio ($DR$) 
of $Y_{\rm A_1,Z_1,H}/Y_{\rm A_1-1,Z_1,H-1}$ to 
$Y_{\rm A_2,Z_2,H}/Y_{\rm A_2-1,Z_2,H-1}$ we obtain from the above formulae 
\begin{equation} \label{yazhdr} 
DR_{A_1A_2}= 
\frac{Y_{\rm A_1,Z_1,H}/Y_{\rm A_1-1,Z_1,H-1}}
     {Y_{\rm A_2,Z_2,H}/Y_{\rm A_2-1,Z_2,H-1}} 
= \alpha_{A_1A_2} 
{\rm exp}\left[-\frac{1}{T}\left(\Delta E_{A_1A_2}^{\rm bh}\right)\right] , 
\end{equation} 
where 
\begin{equation} \label{dEbh} 
\Delta E_{A_1A_2}^{\rm bh}=  E_{A_1}^{\rm bh}-E_{A_2}^{\rm bh} ,
\end{equation} 
and the ratio of the $C$-coefficients we denote as 
\begin{equation} \label{ralph} 
\alpha_{A_1A_2} = C_{\rm A_1,Z_1,H} / C_{\rm A_2,Z_2,H}~. 
\end{equation} 

In central collisions of very high energy leading to production of lightest 
fragments, we can also model a more simple case when the (hyper-) 
fragments are assumed in the final states (i.e., cold ones) in the freeze-out 
volume. 
In such a way we avoid a sophisticated description of the hot fragments, and 
we consider fixed binding energies without a temperature dependence. 
Therefore, within this statistical approach $F_{A,Z,H}$ will be only the 
binding energy of fragments ($E^{b}_{A,Z,H}$) and all above formulae remain 
without modifications but the trivial spin factors. We emphasize that the 
statistical and coalescence interpretation of the data leads to similar 
results in this case \cite{neu03}. 

As one can see from eq.(\ref{yazhdr}), 
the logarithm of the double ratio is directly 
proportional to the difference of the hyperon binding energies in 
$A_1$ and $A_2$ hypernuclei, $\Delta E_{A_1A_2}^{\rm bh}$, divided by 
temperature. Therefore, we can finally rewrite the relation between the 
hypernuclei yield ratios and the hyperon binding energies as
\begin{equation} \label{DRL} 
\Delta E_{A_1A_2}^{\rm bh} = T \cdot \left[ ln(\alpha_{A_1A_2}) - 
ln(DR_{A_1A_2}) \right] .
\end{equation} 
In some cases we expect a large difference in hyperon 
binding energy in both nuclei. For example, according to the liquid-drop 
approach (see eq.~(\ref{fld})) it can be when the difference between $A_1$ and 
$A_2$ is essential (e.g., the mass number $A_2$ is much larger than 
$A_1$). The influence of the pre-exponential $\alpha$ coefficients is small 
and can be directly evaluated, depending on the selected hypernuclei. This 
opens a possibility for the explicit determination of the binding energy 
difference 
from experiments. In this case, it is necessary to measure some number 
of the hypernuclei in one reaction and select the corresponding pairs of 
hypernuclei. One has to identify such hypernuclei, for example, by the 
correlations, and with vertex \cite{star,alice,Vas17,saito-new} or 
'shadow' \cite{Arm93,Ohm97} techniques. However, there is no need to measure 
very precisely the momenta of all particles produced in the reaction 
(including after the week decay of hypernuclei) to obtain their binding 
energy, 
as it must be done in processes of direct capture of hyperons in 
the ground and slightly excited states of the target nuclei (e.g., in 
missing mass experiments \cite{Has06,MAMI}). Therefore, our procedure 
perfectly suits for investigation of 
hypernuclei in the high-energy deep-inelastic hadron and ion induced 
reactions. 

Another interesting way for this study is to use the double ratios of 
yields with the same mass numbers for light and heavy pairs. This case is 
easy to illustrate for cold fragments. The so-called 
strangeness population factor $S$ was introduced in Ref.~\cite{ams04} 
for interpretation of light hypernuclei production in relativistic 
heavy-ion collision (at momenta of 11.5 A GeV/c): 
\begin{equation} \label{pfactor} 
S= 
\frac{Y_{\rm ^3H_{\Lambda}}/Y_{\rm ^3He}} {Y_{\rm \Lambda}/Y_{\rm P}} 
\end{equation} 

Generally, if we involve the pairs of nuclei which differ by one proton 
instead of $\Lambda$-hyperon, we can write the isobar double ratio: 
\begin{equation} \label{ypdr} 
DR^{I}_{A_1A_2}= 
\frac{Y_{\rm A_1,Z_1,H}/Y_{\rm A_1,Z_1+1,H-1}}
     {Y_{\rm A_2,Z_2,H}/Y_{\rm A_2,Z_2+1,H-1}} 
= \alpha^{I}_{A_1A_2} 
{\rm exp}\left[-\frac{1}{T}\left( \Delta E_{X}^{\rm bh} \right)\right] , 
\end{equation} 
where 
\begin{equation} \label{ralph2} 
\alpha^{I}_{A_1A_2} = \frac{g_{\rm A_1,Z_1,H}/g_{\rm A_1,Z_1+1,H-1}}
                     {g_{\rm A_2,Z_2,H}/g_{\rm A_2,Z_2+1,H-1}} , 
\end{equation} 
and the binding energy difference between 4 fragments 
\begin{equation} \label{epdr} 
\Delta E_{X}^{\rm bh}= (E^{b}_{A_1,Z_1,H}-E^{b}_{A_2,Z_2,H})-
(E^{b}_{A_1,Z_1+1,H-1}-E^{b}_{A_2,Z_2+1,H-1}) .
\end{equation} 
The last expression (\ref{epdr}) can not be factorized into the binding 
energies of normal nuclei with $A_1$ and $A_2$ and the part related only 
to the hyperon binding (as it was possible in formula 
(\ref{hybi})), since it includes also the 
difference of the hyperon binding in hyper-nuclei with $Z+1$. Therefore, 
it requires complicated calculations of the coupled equations for extracting 
the hyperon binding. In addition, extra experimental isobar measurements will 
be necessary. Still, the convenient application of $DR^{I}$ can be found for 
single hypernuclei with $H=1$, when for the pair nuclei 
(at $H-1=0$ and $Z+1$) there exist only normal nuclei with known binding 
energies. In this case one can rewrite the formula (\ref{DRL}) as 
\begin{equation} \label{pDRL} 
\Delta E_{A_1A_2}^{\rm bh} = T \cdot \left[ ln(\alpha^{I}_{A_1A_2}) - 
ln(DR^{I}_{A_1A_2}) \right] + \Delta E_{A_1A_2}^{\rm GS} ,
\end{equation} 
where $\Delta E_{A_1A_2}^{\rm GS}$ is the difference of the ground state 
binding energies of non-strange nuclei: 
\begin{equation} \label{edrgs} 
\Delta E_{A_1A_2}^{\rm GS} = (E^{b}_{A_1,Z_1+1}-E^{b}_{A_2,Z_2+1})-
(E^{b}_{A_1-1,Z_1}-E^{b}_{A_2-1,Z_2}) .
\end{equation} 

In the above mentioned example, as 
was obtained by AGS-E864 collaboration \cite{ams04}, $S=0.36$ (with large 
error bars $+-0.26$) for the 
most central collisions and for fragments produced in the midrapidity 
region. The qualitative behavior of this factor with energy was also 
analyzed with dynamical models \cite{Ste12}. 
Since the binding energies of all nuclei in $S$-factor (\ref{pfactor}) are 
known from other experiments we can evaluate from formula (\ref{ypdr}) the 
temperature of the excited hyper-source leading to producing of these 
fragments and hypernuclei: The found chemical temperature is around 
$T\approx 5.5$ MeV. This is typical for the nuclear liquid-gas phase 
coexistence region under condition that all available baryons are produced 
in a dynamical 
way. It is also consistent with the chemical 
temperature and limited equilibration of non-strange fragments 
reported previously for central heavy ion collisions \cite{neu03}. 

%\subsection{Novel directions for application of the method}

It is clear that the suggested 
double ratio approach can be applied to hypernuclei with any number of 
hyperons: Obviously, the equations (\ref{yazh}) and (\ref{yazhdr}) can be 
used for $H > 1$. One can 
reach a multi-strange residues in nuclear reactions with a quite large 
probability \cite{Bot17}, and a very wide mass/isospin range will be 
available for examination. As a result, one can get direct experimental 
evidences for hyperon binding energies in double/triple hypernuclei and 
on influence of the isospin on hyperon interactions in multi-hyperon 
nuclear matter. Such a comprehensive analysis 
is possible within this approach, and it seems the only realistic way to 
address experimentally the hyperon binding in multi-strange nuclei. This 
is an important advantage over the standard hypernuclear measurements. 
Actually, the disintegration of hot hyper-residues suits in the best for 
this examination since all kind of normal and hyper-fragments can be formed 
within the same statistical process. 

The connection between the relative hyperon binding 
energies $\Delta E_{A_1A_2}^{\rm bh}$ and its absolute values can be done 
straightforward: It should be sufficient to make normalization to the binding 
energy of a known hypernuclei (e.g., $A_2$) obtained with other methods. 
However, even relative values are extremely important, when we pursue a goal 
to investigate the trends of the hyperon interaction in different nuclear 
surroundings, e.g., neutron-rich or neutron-poor ones. 
Novel conclusions can be obtained by comparing 
yields of neutron-rich and neutron-poor hypernuclei. The isospin influence 
on the hyperon interaction in matter (revealing in the hyperon binding 
energies) will be possible to extract directly in experiment by using 
the formula (\ref{DRL}). Especially 
multi-strange nuclear systems would be interesting, since they 
can give info on evolution of the hyperon-hyperon interaction depending on 
strangeness. These 
measurements are important for many astrophysical sites, for example, for 
understanding the neutron star structure \cite{Sch08,astro}. 

%\section{Influence of processes accompanying the statistical fragmentation}
%The conception of the statistical formation of fragments in the freeze-out 
%volume suggests, first, the existence of some important parameters (e.g., 
%temperature), and, second, it may suggest the phenomena (e.g., secondary 
%interactions) 
%which can finally change the baryon composition of fragments after they leave 
%the freeze-out. All these effects were under careful examination previously 
%in multifragmentation processes. 

We outline now other details which could be taken into account in the 
hypernuclear case. 
%\subsection{Temperature and the freeze-out state}
In order to find $\Delta E_{A_1A_2}^{\rm bh}$ in experiment within the double 
ratio approach, we should determine the temperature $T$ of the disintegrating 
hypernuclear system. This observable was also under intensive investigation 
recent years in connection with multi-fragment formation. 
There were suggested various methods: using kinetic energies of fragments, 
excited states population, and isotope thermometers \cite{Poc97,Bon98,Kel06}. 
Usually, all 
evaluations give the temperature around 4--6 MeV in the very broad range 
of the excitation energies (at $E^{*} > 2-3$ MeV per nucleon), providing 
so-called a plateau-like behavior of the caloric curve \cite{SMM,Poc97}. 
The isotope thermometer method is the most promising, since it allows for 
involving a large number of normal measured isotopes in the same reactions 
which produce hypernuclei. The corresponding experimental and theoretical 
research were performed last years to investigate better the temperature and 
isospin dependence of the nuclear liquid-gas type phase transition 
\cite{Ogu11,Kel06,Vio01,Buy05}. 
We believe that the great 
experience accumulated previously in this field gives a chance to find a 
reliable temperature of the hypernuclear residues. 

In this case it would be instructive to select the reaction conditions 
leading to similar freeze-out states. The freeze-out restoration methods 
were extensively tested previously: In particular, the masses and 
excitation energies of the hypernuclear residues can be found with a 
sufficient precision \cite{Pie02,Soisson}. 
One can analyze the subsequent ranges of the excitation energy (from low to 
very high ones) to investigate the evolution of the hypernuclei 
with the temperature and the phase transition in hyper-matter. It is specially 
interesting to move into the neutron-rich domain of the nuclear chart, 
by selecting neutron-rich target or projectiles, in addition to sorting out 
the various excitations of the sources. As was previously established in 
multifragmentation studies, the selection of adequate reaction conditions 
can be experimentally verified, 

%\subsection{Secondary de-excitation corrections} 

We may expect that the primary fragments and hyper-fragments (specially, 
large ones) in the freeze-out volume could be excited, therefore, they 
should fastly decay after escaping the freeze-out. For low excited sources 
the fragment excitation energy should roughly correspond to the compound 
nucleus temperature. As was established in theory and multifragmentation 
experiments \cite{hudan}, the internal fragment excitations are around 
2--3 MeV per nucleon for highly excited residue sources. The 
secondary de-excitation influences all 4 fragments entering the double ratio 
and the fragments should loose few nucleons. 
The investigation of similar nuclear decay processes of excited nuclei 
in normal multifragmentation reactions tell us that if the difference in 
mass between initial fragments is small then the mass difference 
between final products will be small too. Following this de-excitation 
the mass numbers will change and we expect a smooth transformation of 
$\Delta E_{A_1A_2}^{\rm bh}$ versus the variation of mass difference 
$\Delta A =(A_2-A_1)$: The new yields and mass numbers should be used for 
the final estimate. This effect can be investigated in 
the framework of the evaporation model for large ($A \goo 16$) hypernuclei 
developed in Ref.~\cite{Bot16}. There was 
demonstrated that mostly neutrons and other light normal particles 
will be emitted from hot large hyper-fragments, since the hyperons have 
a larger binding energy. Such an effect should not change dramatically the 
general form of the $\Delta E_{A_1A_2}^{\rm bh}$ dependence on $\Delta A$. 
For small ($A \loo 16$) hot hyper-fragments formed in the freeze-out volume 
the most adequate model is the Fermi-break-up \cite{SMM}, similar to the 
one for normal fragments. It was generalized for hypernuclei in 
Ref.~\cite{lorente}. 
The consequences of these secondary decay processes will be the subject of 
our future studies. 

%%%%%%%%%%%%%%%%%%%%%%%%%%%%%%%%%%%%%%%%%%%%%%%%%%%%%%%%%%%%%%%%%%%%%%%%
%\begin{figure}[tbh]
%\includegraphics[width=0.6\textwidth]{fig1-hf.eps}
%\caption{\small{ (Color online)
%Probabilities of $^{238}_{\Lambda}$U fission, and ... in competition with 
%evaporation of $\Lambda$ from this hypernucleus 
%versus its excitation energy.
%}}
%\label{fig1}
%\end{figure}
%%%%%%%%%%%%%%%%%%%%%%%%%%%%%%%%%%%%%%%%%%%%%%%%%%%%%%%%%%%%%%%%%%%%%%%%
%
\section{Conclusion}

In conclusion, we should note that during last six decades there is permanent 
increasing the number of 
measured hypernuclei with their binding energies. However, the progress 
is very slow: Because of the special requirements on 
targets in hadron and lepton induced reactions, the traditional hypernuclear 
methods (e.g., the missing mass spectroscopy) can address only a small number 
of isotopes. Also the development of the detectors for measuring nearly 
all produced particles with their exact kinetic energies is very expensive 
and not always practical, that makes problems for a desirable acceleration 
of the studies. 

The suggested double ratio method is related to deep inelastic reactions 
producing all kind of hypernuclei with sufficiently large cross-sections 
in the multifragmentation process. This is a 
typical case for relativistic ion-ion and hadron-ion collisions. Only the 
identification of hypernuclei is required, and, as demonstrated in recent 
ion experiments, there are effective ways to perform it. The experimental 
extraction of the difference in the hyperon binding energies between 
hypernuclei ($\Delta E_{A_1A_2}^{\rm bh}$) is a novel and practical way to 
pursue hypernuclear studies. The advantage of this method over the traditional 
hypernuclear ones is that the 
exact determination of all produced particles parameters (with their decay 
products) is not necessary. Only relative measurements are necessary for 
this purpose, therefore, one can address similar weak-decay chains and 
their products, for example, with the vertex 
technique. The correlation the produced isotopes and particles is an 
adequate information for the double ratios. 

Even more interesting and important that with this method one can also 
determine the difference of hyperon binding energies in double and 
multi-hypernuclei. This gives an access to hyperon-hyperon interactions 
and properties of multi-hyperon matter. It is very difficult to measure 
the hyperon binding energy for exotic (neutron-rich and neutron-poor) nuclear 
species within traditional hypernuclear experiments. On the other hand, 
the hypernuclei with extreme isospin can be easily obtained in 
deep-inelastic reactions. Some of them may have the statistical 
disintegration origin 
and the suggested method opens an effective way for extension of the 
hypernuclear research. 

We believe such kind of research would be possible at the new generation of 
ion accelerators of intermediate energies, as FAIR (Darmstadt), NICA (Dubna), 
and others. It is promising that new advanced experimental installations for 
the fragment detection will be available soon \cite{aumann,frs}. 

A.S. Botvina acknowledges the support of BMBF (Germany) N.B. acknowledges the Turkish Scientific
and Technological Research Council of Turkey (TUBITAK)
support under Project No. 114F328. M.B. and N.B. acknowledges that the work has been performed in the framework of COST Action CA15213 THOR. N.B. thanks the Frankfurt Institute for Advanced Studies (FIAS), J.W. Goethe University, for hospitality during the research visit.

\end{document}